\documentclass{article}

\usepackage{arxiv}

\usepackage[utf8]{inputenc} 
\usepackage[T1]{fontenc}    
\usepackage{hyperref}       
\usepackage{url}            
\usepackage{booktabs}       
\usepackage{amsfonts}       
\usepackage{nicefrac}       
\usepackage{microtype}      
\usepackage{lipsum}
\usepackage{graphicx}
\usepackage[version=3]{mhchem}
\graphicspath{ {./images/} }

\title{Dissolution Dynamics of a Binary Switchable Hydrophilicty Solvent - Polymer Drop into an Acidic Aqueous Phase}

\author{
 Romain Billet \\
  Department of Chemical and Materials Engineering \\
  University of Alberta\\
  Edmonton, T6G 1H9, Canada \\
   \And
Binglin Zeng \\
  Department of Chemical and Materials Engineering \\
  University of Alberta\\
  Edmonton, T6G 1H9, Canada \\
  \And
 James Lockhart \\
  BC Research Inc. \\
  Richmond, BC V6V 1M8, Canada\\
  \And
 Mike Gattrell \\
  BC Research Inc. \\
  Richmond, BC V6V 1M8, Canada\\
  \And
 Hongying Zhao \\
  BC Research Inc. \\
  Richmond, BC V6V 1M8, Canada\\
  \texttt{hzhao@bcri.ca} \\
  \And
 Xuehua Zhang \\
  Department of Chemical and Materials Engineering \\
  University of Alberta\\
  Edmonton, T6G 1H9, Canada \\
  \texttt{xuehua.zhang@ualberta.ca} \\
}

\begin{document}
\maketitle
\begin{abstract}
Switchable hydrophilicity solvents (SHSs) are solvents defined by their ability to switch from their hydrophobic form to a hydrophilic form when put in contact with an acidic trigger such as $CO_2$. As a consequence, SHSs qualify as promising alternatives to volatile organic compounds during the industrial solvent extraction processes, as greener and inexpensive methods can be applied to separate and recover SHSs. Furthermore, because of their less volatile nature, SHSs are less flammable and so increase the safety of a larger-scale extraction process. In this work, we study the dynamics and in-drop phase separation during the dissolution process of a drop composed of SHS and polymer, triggered by an acid in the surrounding aqueous environment. From 70 different experimental conditions, we found a scaling relationship between the drop dissolution time and initial volume with an overall scaling coefficient $\sim 0.53$. We quantitatively assessed and found a shorter dissolution time related with a decrease in pH of the aqueous phase or an increase in initial polymer concentration in the drop. Examining the internal state of the drop during the dissolution revealed an in-drop phase separation behavior, resulting in a porous morphology of the final polymer particle. Our experimental results provide a microscopic view of the SHS dissolution process from droplets, and findings may help design SHS extraction processes for particle formation from emulsions.
\end{abstract}


\section{Introduction}
Solvent removal is one of the most common processes in the chemical industry. Being able to remove and recover the solvent from a system is of the utmost importance and can be done by evaporation induced by heating and/or vacuum. A common way to make such solvent removal and recovery easier is to use low boiling point Volatile Organic Compounds (VOCs). However, these solvents are also usually hazardous pollutants known for their toxicity and smog-forming properties \cite{haagen1952chemistry, atkinson2003atmospheric} These VOCs are also typically highly flammable, requiring special considerations in the design and operation of the solvent recovery processes. Therefore, to respect the principles of green-chemistry, the search for an alternative is necessary. \cite{poliakoff2002green, kerton2013alternative}

Switchable Hydrophilicity Solvents (SHSs) constitute a family of solvents that exhibits hydrophobic (immiscibility with water) properties in their neutral form.  However, when ionized (for example by contact with an acid such as $CO_2$ for a cationic SHS) they become hydrophilic (miscible with water). This switching process is also reversible, and SHS can be "switched" back to their hydrophobic form by, for example, removing $CO_2$ by flushing with $N_2$. \cite{jessop2010solvent} \cite{vanderveen2014design}  The mechanism behind this switching process is the acid-base chemical reaction described below:

 \[\ce{SHS_{org} + H^+_{aq}  <=> SHSH^+_{aq}} \]
 
 For the $CO_2$ trigger, it is the dissolution of $CO_2$ and the acidification of water that allows the SHS to switch. This switching behavior can more generally be observed with other kinds of acid. \cite{shahvandi2018development, shahraki2018electrochemical, oenning2020green}

 SHSs present an economically viable and green alternative to VOCs due to the simplicity to "switch on" and "off" the hydrophobicity of the solvent with high recoverability coming from the "switching off" process. Potential applications of SHS have already been demonstrated in soybean oil extraction\cite{phan2009soybean} \cite{viner2019transesterification}, separation of bitumen from oil sands\cite{holland2012separation}, biofuel extraction from microalgae  \cite{boyd2012switchable, du2015opportunities}, polystyrene foam \cite{jessop2011tertiary} and multi-layer packaging recycling\cite{samori2017application, mumladze2018sustainable}, liquid-liquid micro-extraction (SHS-LLME) for analytical chemistry  \cite{alshana2020switchable,bazel2020switchable} or latex formation
 \cite{su2017preparing}.

In latex formation, binary drops of SHS and dissolved polymer are made to deposit solid polymer particles by switching off SHS hydrophobicity in an aqueous environment. A typical example is the formation of polystyrene particles by the dissolution of N-N Dimethylcyclohexylamine (DMCHA) from binary drops of DMCHA and polystyrene. When the DMCHA-polystyrene drop contacts an acidic phase, DMCHA reacts with protons producing the protonated counterpart DMCHAH+ following the biphasic reaction below:

 \[\ce{DMCHA_{org} + H^+_{aq}  <=>> DMCHAH^+_{aq}} \]
 
The produced DMCHAH$^{+}$ is solubilized in the aqueous environment at the surface of the binary drop. In this way, the DMCHA in the DMCHA-polystyrene drop begins to dissolve into the aqueous phase and the drop shrinks with time. Subsequently, the insoluble polystyrene part of the drop is left behind and forms a solid polystyrene particle. Understanding of the dissolution dynamics of the binary drop is required to obtain the desirable morphology, size, and properties of final polymer particles.

The dissolution process of drops with a low solubility in their environment is a problem that has already been studied previously. In an early work, Duncan et al \cite{duncan2006microdroplet} showed experimentally that the diffusion-driven model for bulk bubbles in an undersaturated liquid phase developed by Epstein \cite{epstein1950stability} could be also applied to dissolution of a free oil microdroplet.  The diffusion-driven model was further developed to describe the dissolution of a drop deposited on a substrate (i.e. sessile drop) with effects of the drop geometry taken into account. \cite{popov2005evaporative} The lifetime of a dissolving sessile drop in various modes such as constant contact radius, constant contact angle, stick-slide, or stick-jump modes of the drop have been experimentally and theoretically studied.\cite{zhang2015mixed, dietrich2015stick, bao2018flow}

To add to the diffusion-driven dissolution, the influence of gravity-induced convection in the bulk liquid on the dissolution process has also been shown to speed up the dissolution dynamics.\cite{dietrich2016role}. More recently, the dissolution process of multicomponent drops, as opposed to pure liquid drops, has been found to exhibit complex dynamics. The components in the drop may undergo preferential dissolution \cite{tan_diddens_mohammed_li_versluis_zhang_lohse_2019}, phase separation \cite{dietrich2017segregation}, or self-assembling \cite{lu2017dissolution}. For instance, dissolution of polymer solution drops may lead to formation of polymer capsules \cite{sharratt2018microfluidic, udoh2019polymer, watanabe2014microfluidic, jativa2017transparent}, while snowballs of graphene oxide may develop from dissolution of colloidal drops. \cite{yang2012assembling} \cite{yang2015tailoring}

The dynamic of the switching SHS and corresponding solvent extraction process has previously been shown to be time consuming, in some cases reaching hours \cite{vanderveen2014design}. It is therefore important to understand what conditions may possibly shorten the switching-extraction time and minimize residual solvent in a cost-effective manner. In the extraction process assisted by $CO_2$-switching SHS, two steps may play a role: (1) the slow mass transfer of $CO_2$ gas into the aqueous phase, followed by the chemical reactions of dissociation and hydrolysis of $CO_2$ to $HCO_3^{-}$, $CO_3^{2-}$ and $H^+$, and (2) the liquid phase reaction (switching) of neutral SHS (hydrophobic form) and dissolution of switched SHS (hydrophilic form) into the aqueous solution. In previous work by Han et al the acceleration of the switching-extraction dynamics was studied using a microfluidic device. The improved specific interfacial area between the aqueous phase and $CO_2$ gas accelerated the extraction process inside the microfluidic device. \cite{han2021intensified,han2020accelerated} However, to the best of our knowledge, there is no quantitative understanding on the SHS dissolution dynamics of the reaction-induced mass transfer of the SHS, and in-particular inside a liquid drop and its impact on the final morphology of the polymer particles post solvent extraction. 

In this work, we study the dissolution of a binary DMCHA/polystyrene sessile drop immersed in a controlled acidic aqueous solution environment as it is presented in Fig. \ref{sketch}. We aim at understanding the reaction-induced mass transfer of the SHS from the drop to the aqueous phase. We find a scaling relationship between the drop dissolution time and the initial size of the drop, and study the impact of the aqueous phase pH and initial drop composition on this scaling law. We also follow the internal state of the drop during the dissolution process, and show the existence of a phase separation behavior and its implication on the final morphology of the final polymer particle. Our findings may help to give a quantitative understanding of the switching dynamics of SHS drops during the initial external reaction/diffusion dominated phase, and provide useful insights for internal drop dynamics and improved design for SHS switching processes in applications such as latex formation.

\section{Experimental section}

\subsection{Materials, solutions and substrates}
The SHS polymer solutions were prepared using a mixture of N-N Dimethylcyclohexylamine (Sigma-Aldrich, 99\%, DMCHA) and polystyrene (Sigma-Aldrich, $M_{w}$ = 40 000 g/mol, beads >99\%). The weight ratio ranged from 10 to 30 wt\% of polystyrene. Both chemicals were mixed under ambient condition in sealed vials using a magnetic stirrer for 2 hours until complete dissolution of the polystyrene beads in DMCHA. Acid solutions were freshly made before each dissolution using formic acid (Sigma-Aldrich, > 95\%) and (Milli-Q) water. The solution concentration used in this work ranged from $1.1 \times 10^{-1}$ M to $2.7 \times 10^{-4}$ M (0.4 vol\% to 0.001 vol\%). The pH of the solutions was measured using a benchtop pH meter (Fisher Scientific, Accumet AE150) and ranged from 2.37 to 3.83.  Small square ($1 \times 1 \ cm$) silicon substrates were cut from wafers and thoroughly rinsed with water, ethanol and then sonicated for 20 minutes with ethanol before finally being dried with a stream of air. The partial solubilities of DMCHA in water as well as water inside of DMCHA were measured by progressively adding one compound in a vial of the other until saturation. A solubility of DMCHA in water of $15.3 \pm 1.7 \ g/L$ and water in DMCHA of $155 \pm 7 \ g/L$ ($18.3 \pm 0.8$ vol$\%$) were measured close to what was previously reported in the literature. \cite{stephenson1993mutual}

\subsection{Drop dissolution}

\begin{figure}
 \centering
 \includegraphics[width=16cm]{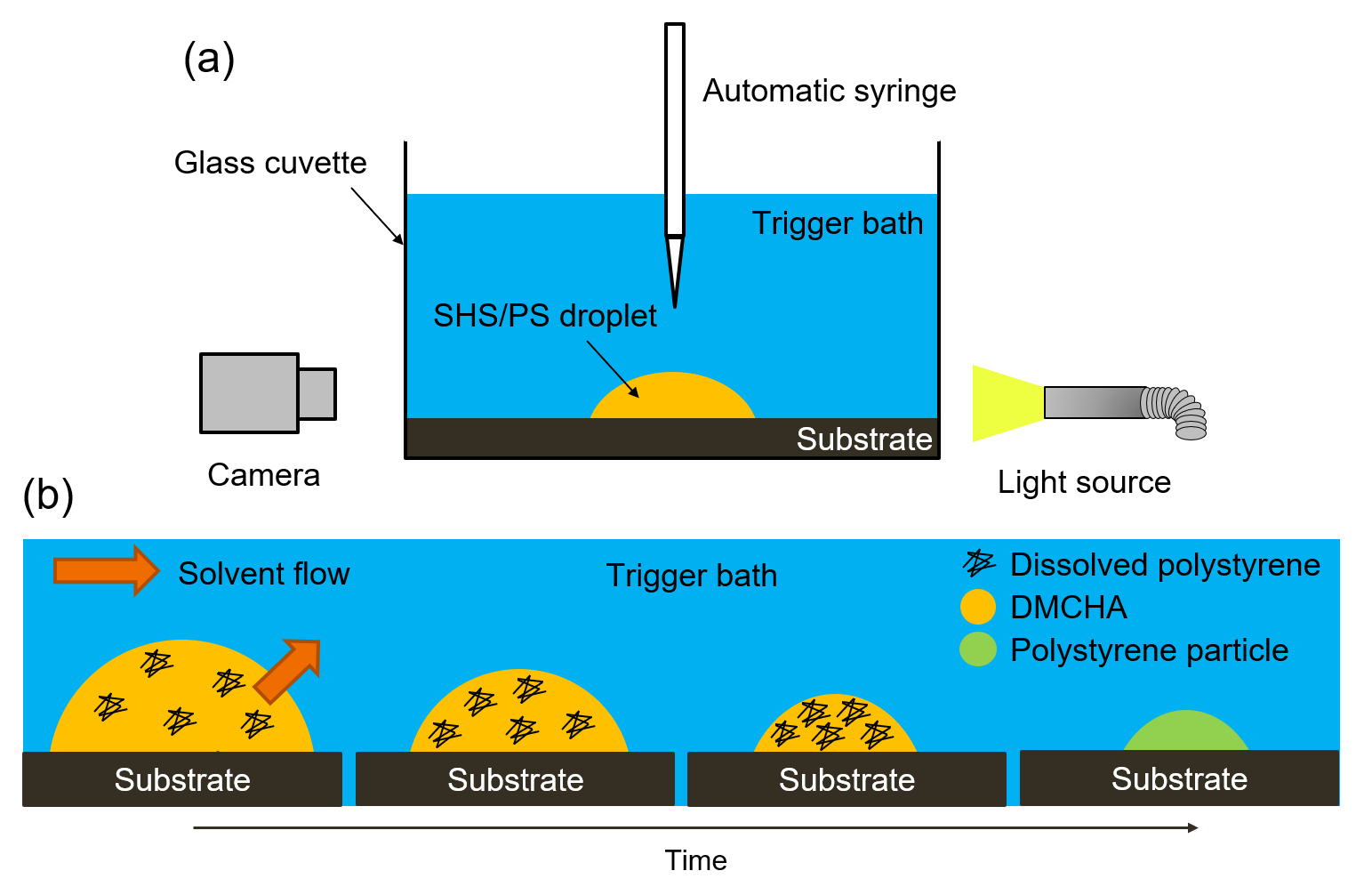}
 \caption{(a) Sketch of the single drop dissolution experimental setup. The binary DMCHA/polystyrene drop is placed on a silicon substrate inside a cuvette filled with FA solution using a motorized automatic syringe. The side view imaging is done with the help of a side camera and a light source passing through the drop. (b) Schematic depiction of the drop dissolution process. DMCHA is extracted out of the drop until only a polystyrene particle remains. }
 \label{sketch}
\end{figure}

To study the dynamics of this dissolution, a side view camera was used to record the drop dissolution. The drop dissolution setup is shown in Fig. \ref{sketch}. The substrates were placed at the bottom of a glass cuvette (Krüss Scientific, $36\times36\times30 \ mm \ W\times D\times H$). The polymer solution was then filled in a glass syringe that was controlled by a connected Drop Shape Analyzer instrument (DSA-100, Krüss Scientific).

For each experiment, the glass cuvette was filled with 30 mL of the acidic solution and drops of initial volume ranging from 0.15 to 3.65 $\mu$L were introduced by the motorized syringe on the surface of the substrate inside the aqueous acidic phase. In total, 70 dissolution conditions were studied, as summarized in Table \ref{tablePS} for the polystyrene content parameter and in Table \ref{tablepH} for the pH levels. During the drop dissolution, the side view images were captured directly by the DSA-100 with a 9x magnification. For the top view, an upright optical microscope (Nikon H600l) equipped with a 4x magnification lens and a camera was used to image the drops.

After the drop dissolution, the substrate was carefully removed from the acidic solution. The excess water on the sample was gently blown with air and the remaining water left to dry under ambient conditions for 48 hours.

\begin{table*}

\small
 \caption{Experimental parameters to study the influence of the initial drop composition, the pH in the aqueous phase is fixed at 2.53.}

\label{tablePS}

\begin{tabular}{|l|l|l|l|l|l|l|l|l|l|l|l|l|l|l|l|} 
\hline
Initial polystyrene (wt\%) & \multicolumn{9}{c|}{0}                                            & \multicolumn{4}{c|}{10}\\ 
\hline
Drop volume ($\mu L$)                & 0.15 & 0.20 & 0.35 & 0.50 & 0.55 & 0.65 & 0.80 & 2.30 & 4.25 & 0.25 & 0.40 & 0.55 & 0.70  \\ 
\hline
\end{tabular}

\begin{tabular}{|l|l|l|l|l|l|l|l|l|l|l|l|l|l|l|l|} 
\hline
Initial polystyrene (wt\%) & \multicolumn{6}{c|}{10}                                            & \multicolumn{7}{c|}{20}                                                          \\ 
\hline
Drop volume ($\mu L$)            & 0.90 & 1.20 & 2.00 & 2.40 & 2.95 & 3.65 & 0.30 & 0.45 & 0.70 & 0.85 & 0.95 & 1.45 & 2.20  \\ 
\hline
\end{tabular}
\begin{tabular}{|l|l|l|l|l|l|l|l|l|l|l|l|l|l|l|l|l|} 
\hline
Initial polystyrene (wt\%)                                          & \multicolumn{3}{c|}{20}  & \multicolumn{9}{c|}{30}                                                 \\ 
\hline
Drop volume ($\mu L$)               & 2.55 & 3.00 & 3.25 & 0.35 & 0.75 & 1.00 & 1.30 & 1.50 & 2.10 & 2.40 & 2.75 & 2.85  \\
\hline
\end{tabular}

\end{table*}

\begin{table*}

\caption{Experimental parameters to study the influence of the trigger concentration, the initial polystyrene composition of the drop is fixed at 10 wt\% }

\label{tablepH}
\small

\begin{tabular}{|l|l|l|l|l|l|l|l|l|l|l|l|l|l|l|l|l|l|} 
\hline
Aqueous phase pH & \multicolumn{5}{c|}{3.83}    & \multicolumn{9}{c|}{3.05}                                    \\ 
\hline
Drop volume ($\mu L$)   & 0.20 & 0.60 & 1.35 & 1.70 & 2.60 & 0.15 & 0.40 & 0.45 & 0.75 & 1.10 & 1.40 & 1.75 & 2.00 & 2.85  \\
\hline
\end{tabular}

\begin{tabular}{|l|l|l|l|l|l|l|l|l|l|l|l|l|l|l|l|l|} 
\hline
Aqueous phase pH &  \multicolumn{1}{c|}{3.05}  & \multicolumn{10}{c|}{2.53}                                  & \multicolumn{3}{c|}{2.37}  \\
\hline
Drop volume ($\mu L$) & 3.20 & 0.25 & 0.40 & 0.55 & 0.70 & 0.90 & 1.20 & 2.00 & 2.40 & 2.95 & 3.65 & 0.15 & 0.60 & 1.20    \\
\hline
\end{tabular}

\begin{tabular}{|l|l|l|l|l|l|l|l|l|l|l|l|l|l|l|l|l|} 
\hline
Aqueous phase pH &  \multicolumn{3}{c|}{2.37}  \\
\hline
Drop volume ($\mu L$)  & 1.80 & 2.20 & 2.85  \\
\hline
\end{tabular}

\end{table*}

\subsection{Image analysis and characterization}
Video footage of the drops was analyzed using MATLAB. The MATLAB code used analyzed each frame of the side view videos and detected the drop using an intensity threshold. The detected drop was then used to measure the quantities of interest. This can be done either by assuming a spherical model and fitting a spherical-cap shape to the contour of the drop, which allows access to the contact angle of the drop, or by the 180° symmetric rotation of the drop section observed to obtain a volume. However, at the end of the dissolution process, the drop deviates from the spherical model due to the pinning and deformation of the drop. Therefore, we use the rotation to calculate an approximation of the drop volume.  The final particles micro-structure and surface state were characterized by SEM (Zeiss Sigma FESEM).  A theoretical final to initial volume ratio was calculated assuming (i) the complete removal of DMCHA from the particle, (ii) the absence of any porosity, and (iii) by approximating the DMCHA/polystyrene mixture density with a combination of the density of its pure constituents $\frac{1}{\rho_{mix}} = \frac{x_{DMCHA}}{\rho_{DMCHA}} + \frac{x_{PS}}{\rho_{PS}}$, with $\rho_{mix}$ the DMCHA/polystyrene solution density, $\rho_{DMCHA}$ and $\rho_{PS}$ respectively the pure DMCHA and polystyrene density, and $x_{DMCHA}$ and $x_{PS}$ respectively the mass fraction of DMCHA and polystyrene.

\section{Results}

\subsection{Morphology of a dissolving binary SHS/polymer drop}

\begin{figure*}
 \centering
 \includegraphics[width=16.5cm]{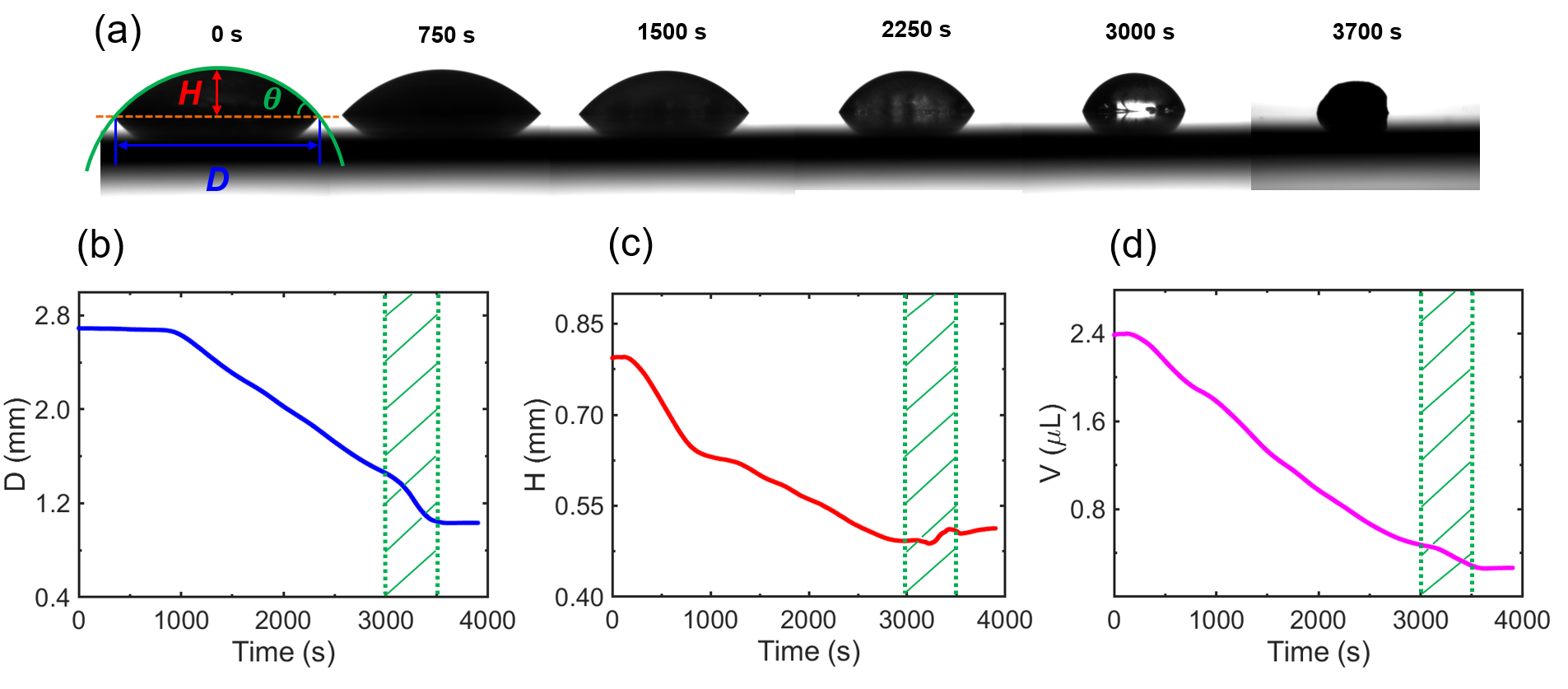}
 \caption{(a) Snapshots of a 2.4 $\mu L$, 90:10 wt\% DMCHA:polystyrene drop inside of an aqueous acid solution of pH 2.53 at different times during the dissolution process. The dashed orange line corresponds to the contact line between the drop and the substrate. The curved green line shows the fitting of the drop to a spherical model. H is defined as the height of the drop, D shows the contact diameter and $\theta$ is the contact angle of the drop.  (b-d) Plots of different parameters of the drop (Contact diameter (b), height (c) and volume (d)) measured and calculated via image analysis against the dissolution time. The green dashed zone corresponds to the observed morphology deformation phase when the liquid drop transitions to a solid particle. }
 \label{observation}
\end{figure*}

The dissolution process was recorded and then the videos were converted into image frames vs time. Fig. \ref{observation}(a) presents a typical dissolving process of a drop. The initial composition in the drop was 90:10 wt\% DMCHA:polystyrene dissolved in a FA solution of pH = 2.53. In the first snapshot, we show the baseline which corresponds to the contact line between the drop and substrate. Using a spherical-cap shape model, we fit a circle to the rim of the drop, allowing to define several quantities to describe the drop state. D is defined as the contact diameter and H as the height of the drop. We also define $\theta$ as the contact angle, the tangential angle of the circle at the baseline intersection. The time t = 0 is taken at the start of the recording, which corresponds roughly to the time of the drop deposition with a time difference of around 5s, negligible compared to the total time of dissolution $\sim$ 3,000s. At the beginning of the dissolution, the drop exhibits a spherical-cap shape that allows to clearly define the contact angle (t = 0 - 3,000s). However, approaching the end of the dissolution (t = 3,000 - 3,500s), deformation starts to appear and the drop deviates from the spherical-cap shape model. Therefore, throughout this step, the measurement of a contact angle becomes less accurate. The morphology deformation may also correspond to the shift from the liquid drop to the solid particle and so the contact angle at this time would no longer have a meaning related to the liquid properties. Following this, the dissolution ends and no further changes can be observed.  During the dissolution process, the drop contact diameter shrinks from 2.7 mm to 1 mm. However, the contact angle increases by 30° between the initial and final state. 

Fig. \ref{observation}(b-c) shows the measured contact diameter $D$ and drop height $H$ against the dissolution time $t$.  The contact diameter remains constant at D = 2.7 mm from t = 0 to t = 1,000s as the drop boundary is pinned on the surface. Then, the contact diameter decreases during the remaining of the process (1,000s < t < 3,500s).  The height of the drop decreases from 0.8 mm to 0.5 mm during almost the entire process. At first, we observe a sharp drop of H  during the constant contact diameter step from t = 0 to t = 1,000s, and then a slow decrease from t = 1,000 s to t = 3,000s. Afterward, morphology change of the drop sometimes induces a slight height increase.  

Fig. \ref{observation}(d) shows the change in calculated volume V of the drop. As we only have snapshots of a side view, we assume a 180° rotational symmetry of the drop around its center. We can therefore calculate the drop volume by measuring only the surface area of the observed drop section. This hypothesis is confirmed by the top view imaging, and only deviates slightly from the symmetric model during the deformation step.  Throughout the entire dissolution process, the calculated volume of the drop decreases from 2.4 $\mu L$ to 0.3 $\mu L$.  The final volume is approximately 12.5\% of the initial volume, which is close to the calculated ratio of 12.2\% based on the initial drop volume and polymer mass fraction.

\subsection{Dependence of dissolution on initial drop volume}

\begin{figure*}
 \centering
 \includegraphics[height=8cm]{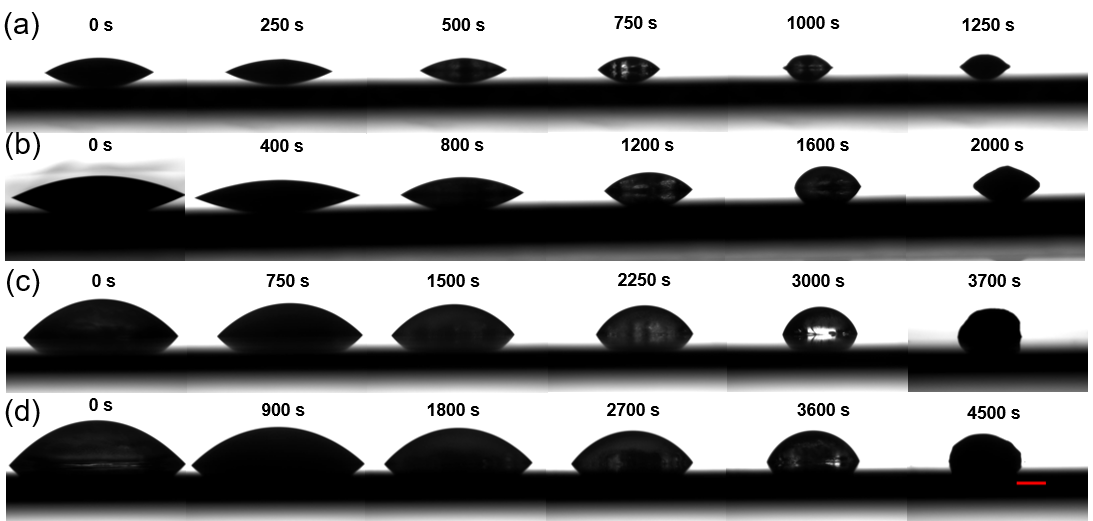}
 \caption{(a-d) Snapshots during the dissolution process of different initial drop volume (from (a) to (d) 0.25, 0.90, 2.40 and 2.95 $\mu L$). The drops all have a 90:10 DMCHA:polystyrene initial composition and are dissolving inside of an aqueous acidic environment of pH = 2.53. Smaller drops are dissolving faster than larger drops. The length of the scale bar is 0.5 mm.}
 \label{initial volume}
\end{figure*}

A series of dissolution experiment was performed to study the influence of different parameters on the drop shrinking dynamics, including the polystyrene content and acidic solution pH. The first analyzed parameter is the initial drop volume. The drop volumes range from 0.25 $\mu L$ to 3.65 $\mu L$. Fig. \ref{initial volume}(a-d) shows snapshots of the dissolution of 90:10 wt\% DMCHA:polystyrene drops in a FA solution of pH = 2.53 with different initial volumes. As expected, the lifetime of smaller drops is shorter than that of bigger drops. The timescale of the dissolution can be as short as 950s ($\sim 16$ min) for the smallest drops (0.25 $\mu L$) and as long as 4700s ($\sim 78$ min) for the largest drops (3.65 $\mu L$).

To quantify the dissolution process dynamics, we introduce a scaling law. The initial volume of the drop $V_0$ is defined as the measured volume at the time t = 0. We also define the lifetime $\tau$ of the drop, which corresponds to the time when the measured volume V becomes constant as a function of time, and correlates to the time when the dissolution is finished and no changes can be observed anymore (less than 0.01\% of volume change per second for a minimum of 30s). We relate $\tau$ and $V_0$ with the following scaling law:

 \begin{equation}
 \tau = \alpha * V_0^{\gamma} 
  \end{equation}
 
 \par The coefficient $\alpha$ is a prefactor whereas $\gamma$ is a scaling coefficient. 
 
\begin{figure*}
 \centering
 \includegraphics[height=6.4cm]{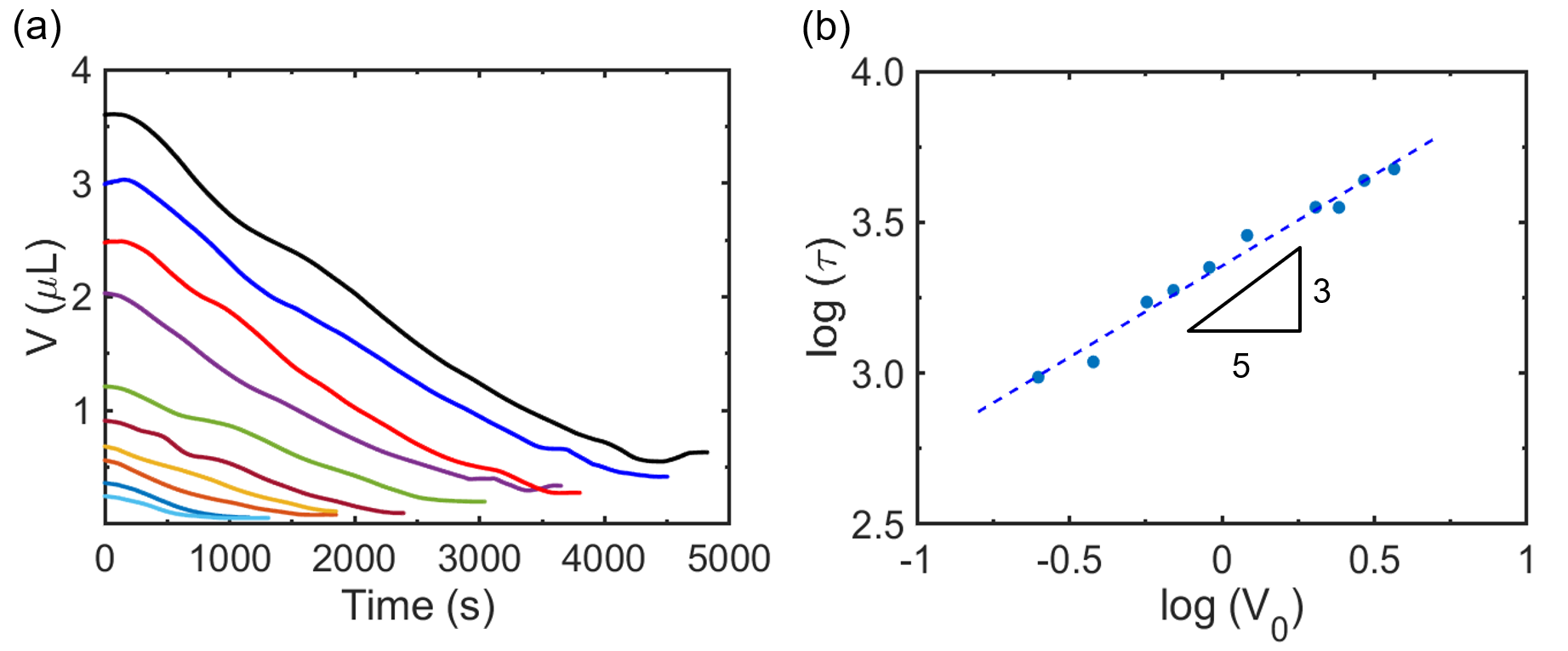}
 \caption{(a) Calculated volume as a function of the time for drops with different initial volume $V_0$. The drops all have a 90:10 DMCHA:polystyrene initial composition and are dissolving inside of an aqueous acidic environment of pH = 2.53. (b) Scaling law as a log-log plot of the drop lifetime $\tau$ as a function of the initial volume $V_0$. Points are extracted from (a) and a line of equation $\gamma x + b$ is fitted to the experimental data. The slope is the scaling coefficient $\gamma$ and the intercept $b$ is related to the prefactor with $\alpha = 10^b$ }
 \label{fgr:scaling law}
\end{figure*}

 
Fig. \ref{fgr:scaling law}(a) shows the volume curves as a function of the dissolution time. We then extract from the volume curves the drop initial volume $V_0$ and lifetime $\tau$ which are then used to deduce the scaling law presented in Fig. \ref{fgr:scaling law}(b). A scaling coefficient $\gamma$ of $ \sim 0.60 $ and prefactor $\alpha$ of $ \sim 2200$ are found from the curve fitting to our scaling model.

\subsection{Influence of the initial drop composition and trigger concentration}

We study the impact of the initial drop composition and the trigger concentration on the dissolution process by fixing one parameter and changing the other.  Video footages of the dissolution of 6 to 12 drops of different initial volume are analyzed.

\par
\begin{figure*}
 \centering
 \includegraphics[height=6.4cm]{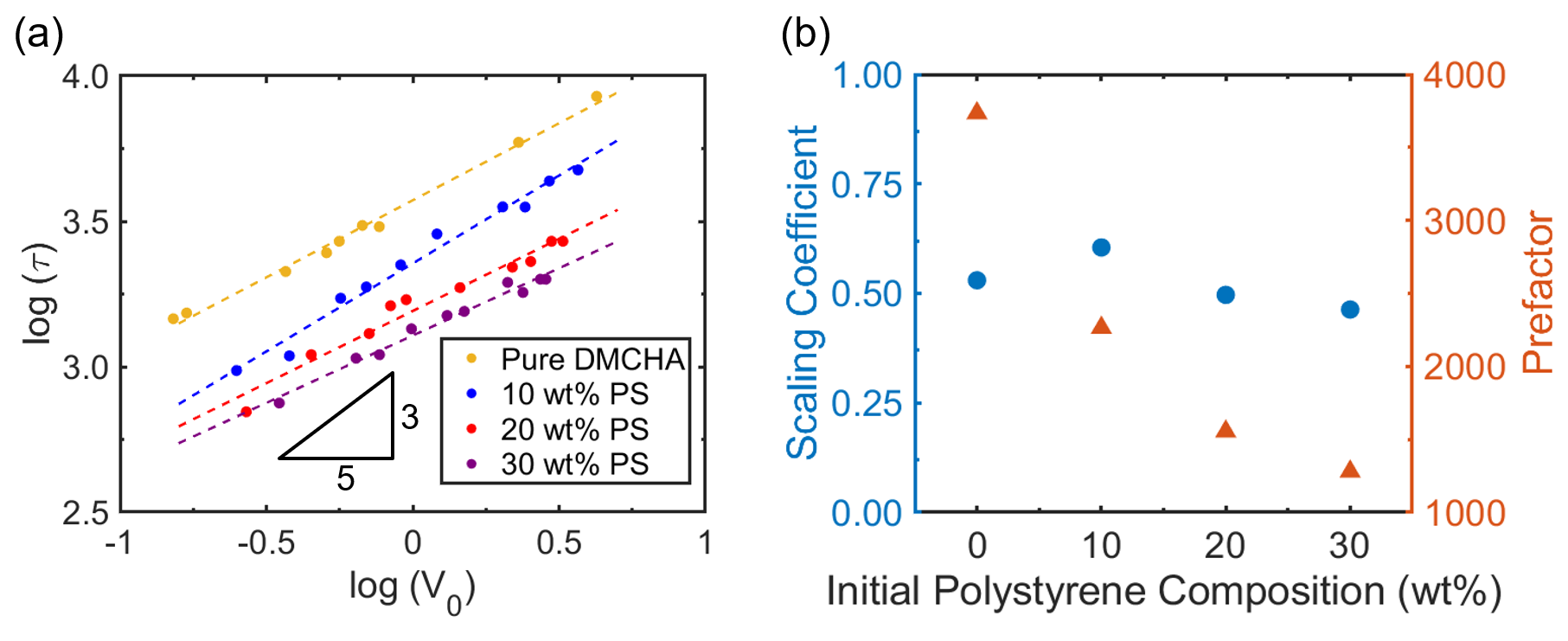}
 \caption{(a) Scaling laws as a log-log plot of the drop lifetime $\tau$ as a function of the initial volume $V_0$ for different initial drop composition. The polymer solutions used are from top to bottom, pure DMCHA, 10, 20 and 30 wt\% of PS. Every point represent one drop dissolution. (b) Scaling coefficients (left) and prefactors (right)  extracted from the scaling laws as a function of the initial polystyrene composition in wt\%.}
 \label{composition}
\end{figure*}

 Firstly, we fix the trigger concentration to study the impact of the initial drop composition. The pH of the aqueous solution is fixed at 2.53, while initial drops contain 10, 20 and 30 wt\% of polystyrene. The plot of $\tau$ against $V_0$ is presented Fig. \ref{composition}(a). We can see that the fitted lines for different polystyrene concentrations in the drop are almost parallel, suggesting similar scaling coefficient $\gamma$, as it is summarized in Fig. \ref{composition}(b). The scaling coefficient seems to be constant against the change in polystyrene initial concentration, and a mean scaling coefficient $\gamma = 0.52 \pm 0.07$ is found.  However, the line fittings differ in the prefactor as shown in Fig. \ref{composition}(b). A trend is that the prefactor decreases with an increase in polystyrene initial concentration. This physically means that the higher the initial polystyrene concentration is, the faster the same-sized drop dissolution will finish.

\par
\begin{figure*}
 \centering
 \includegraphics[height=6.4cm]{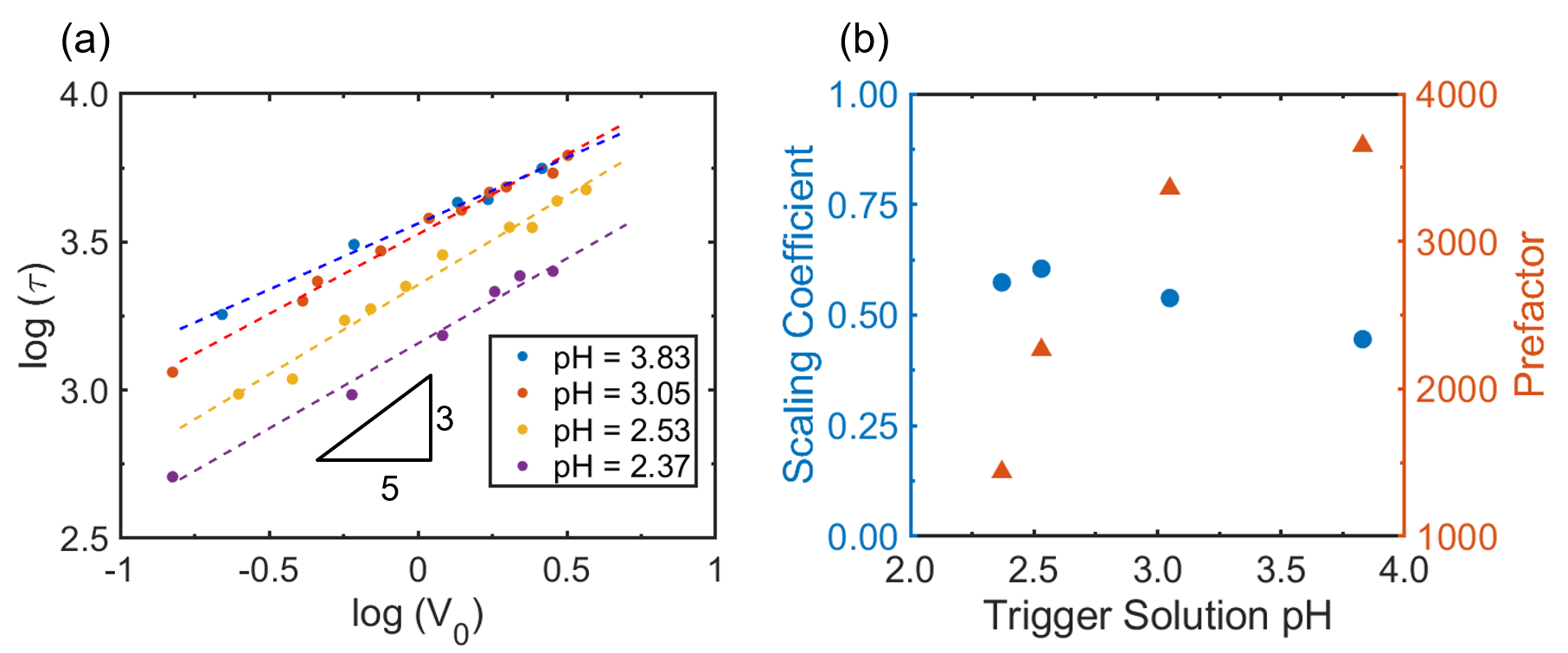}
 \caption{(a) Scaling laws as a log-log plot of the drop lifetime $\tau$ as a function of the initial volume $V_0$ for aqueous solution pH. From top to bottom, the pH of the aqueous extracting phase is 3.83, 3.05, 2.53 and 2.37. Every point represents one drop dissolution. (b) Scaling coefficients (left) and prefactors (right)  extracted from the scaling laws as a function of the aqueous solution pH.}
 \label{concentration}
\end{figure*}

 We also study the impact of the trigger concentration by fixing the initial drop composition. During the experiments, the polymer solution concentration used to make the drop is maintained at 10 wt\% of polystyrene. We change the trigger concentration inside of the aqueous phase from 0.001 vol\% of FA (pH = 3.83) to 0.4 vol\% of FA (pH = 2.37). The resulting $\tau$ against $V_0$ is presented Fig. \ref{concentration}(a). Again, we can see that the fitted lines are nearly parallel. We observe an almost constant scaling coefficient in Fig. \ref{concentration}(b). The mean scaling coefficient $\gamma$ is $\gamma = 0.54 \pm 0.07$. Similarly to the drop initial composition parameter, the fittings differ by their prefactor, as summarized in \ref{concentration}(b). A higher trigger or formic acid concentration results in a lower pH and lower prefactor. Therefore, at lower pH, the dissolution process proceeds faster than at higher pH.

\subsection{Internal view of the dissolution process}

\begin{figure*}
 \centering
 \includegraphics[width=16cm]{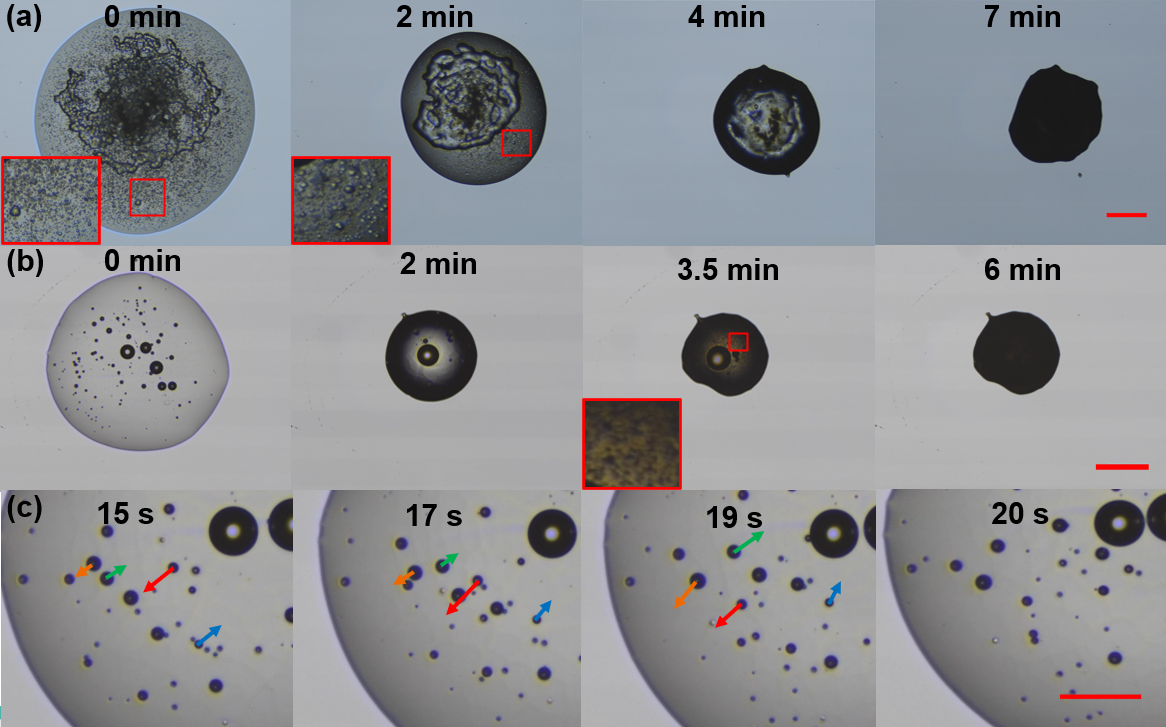}
 \caption{(a) Snapshots of a 90:10 wt\% DMCHA:polystyrene drop observed from the top in an aqueous phase at pH 2.53 at different time during the dissolution. The drop internal state can be observed. At the center of the drop, a polystyrene precipitate is present and microdroplets can be seen moving along the radius of the principal drop. The scale bar corresponds to 200 $\mu m$ (b) Snapshots of a DMCHA:polystyrene:Water drop at saturation observed from the top in an aqueous phase at pH 2.53 at different time during the dissolution. No polystyrene precipitates can be observed, however microdroplets of water can be seen moving along the radius of the main drop and coalescing during the dissolution. The scale bar corresponds to 200 $\mu m$ (c) Zoom into the main water and polystyrene saturated drop. The water microdroplets move radially inside of the drop, going either to the rim or the center of the drop. The scale bar corresponds to 60 $\mu m$}
 \label{top view}
\end{figure*}

To complete the dissolution process observation, we setup a top view camera. This top view experiment allows us to discern the internal structure and flow in the drop during the extraction of DMCHA. Fig \ref{top view}(a) presents snapshots at different stages during the dissolution of a 90:10 DMCHA:polystyrene drop inside of an acidic solution of pH 2.53.

When the drop is introduced in the trigger bath, multiple phases can be observed. Once the dissolution starts, three phases are observed within the drop: a solid polystyrene phase at the center of the drop, small microdroplets moving radially inside of the liquid phaseof polymer and unswitched DMCHA. Such phase separation behavior appears during the early stage of the dissolution process and can be considered as almost instantaneous compared to the remainder of the drop shrinking.

\par As time passes, the main drop shrinks in accordance to what was observed with the side-view experiment. Some of the microdroplets seen inside of the main drop join with the polystyrene aggregate in the center, leading to a change in the shape of the polystyrene core. Finally, the liquid outline of the main drop touches the edge of the polystyrene aggregate (t = 4min). At this time, deformation of the drop starts to occur, and the main drop boundary starts to coincide with the polystyrene core. During the remaining solvent dissolution, the drop finishes its deformation step resulting in the formation of the final polystyrene particle.

\par As we suspected the aqueous environment to saturate the drop with water and alter the solubility of polystyrene, we also examined a solution of DMCHA/polystyrene made from DMCHA pre-saturated with water. Water pre-saturated DMCHA did exhibit a much lower polystyrene solubility qnd wqs brought to saturation with polystyrene. During the mixing of water-saturated DMCHA with polystyrene, water microdroplets also separate from DMCHA and the system becomes an emulsion of water inside of the DMCHA/polystyrene solution. The emulsion is stable and takes as long as two days to separate. The saturated DMCHA drop dissolution is presented Fig \ref{top view}(b). The drop pre-saturated with water and polystyrene, as opposed to the one without pre-saturation with water does not exhibit any internally precipitated polystyrene at the beginning of the dissolution. Water microdroplets coming from the emulsion range from 5 to 50 $\mu m$ in diameter. During the shrinking, the microdroplets can be seen moving radially. The microdroplets also grow larger as they are coalescing with each other. Near the end of the dissolution process, the main drop starts to deform and at this step, a large number of small-scale microdroplets measuring less than 1 $\mu m$ can be seen appearing Fig. \ref{top view}(b) (t = 3.5 min). Some of the water droplets appear stay trapped in the particle as the dissolution process finishes.

\begin{figure}
 \centering
 \includegraphics[width=16cm]{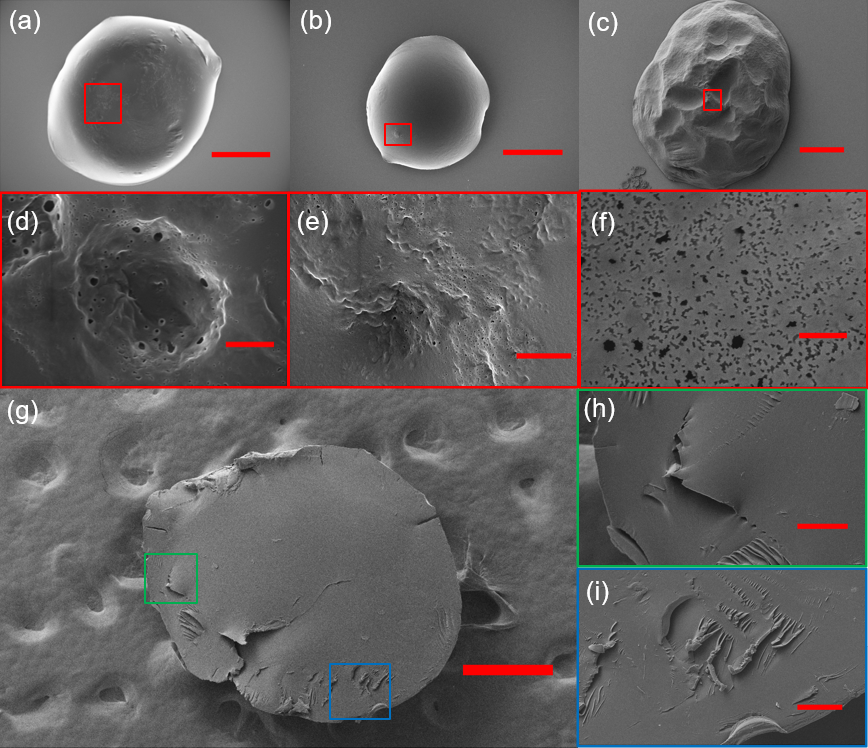}
 \caption{(a-c) SEM images of the final particle obtained after the drop dissolution process. From left to right, different conditions of drop dissolution have been applied, respectively, (a) 90:10 wt\% DMCHA:polystyrene at pH 2.53, (b) 80:20 wt\%, DMCHA:polystyrene at pH 2.53 and (c) 90:10 wt\% DMCHA:polystyrene at pH 2.37 . Scale bars: 400 $\mu m$, 400 $\mu m$ and 200 $\mu m$ (d-f) Enlargment of the microstructure of the particles. A porous surface can be observed at the micrometer scale. Scale bars: 800 $nm$, 600 $nm$ and 800 $nm$  (g) Bottom surface of the particle previously in contact with the silicon substrate. The particle was produced with a 90:10 wt\% DMCHA:polystyrene drop at pH 2.53. The particle is turned upside down. Enlargments of cracks (h) and wrinkles (i) observed at the rim of the particle are presented respectively in green and blue. Scale bars: 400 $\mu m$, 60 $\mu m$ and 60 $\mu m$}
 \label{structure}
\end{figure}

\subsection{Porous microstructure of the final particles}

We also investigate the final surface state and microstructure of the particles resulting from the dissolution process using SEM. Multiple particles were prepared using drop composition and trigger concentration as parameters. Fig. \ref{structure}(a-f) present SEM images of 3 particles. We also examine the bottom surface of the particle, which corresponds to the portion of the particle in contact with the silicon substrate Fig. \ref{structure}(g).

\par Firstly, the overall shape of the structure exhibit an almost spherical shape, even while changing the conditions of the dissolution. A slight pinning effect of the substrate is observed in the final shape of some of the particles. However, a difference in the surface state can be observed when the pH of the solution used is increased. For a high pH aqueous solution, the final particles exhibit a rougher surface at a macroscopic scale, while at low pH, the surface state seems to be smoother.

\par Another characteristic of the particles can be observed when looking at the microstructure of the particle. Some pores are observed at the micro-scale, with pores diameters ranging from 30 $nm$ to 200 $nm$. We observe larger pores in structures starting with a lower polystyrene initial composition, resulting in smoother surfaces. On the opposite side, smaller pores are seen for higher polystyrene initial composition drops.

\par The bottom surface of the particle presented in Fig. \ref{structure}(g) exhibits a smooth texture without pores or defects for the area around the center of the particle. However, at the rim of the particle, different defects such as cracks or wrinkles are observed. The size of the defects ranges from 10 $\mu m$ to 100 $\mu m$. Once more, in the rim area of the particle, no pores can be observed when looking at micrometer scale.

\section{Discussion}

\subsection{A switching-dissolution process}



An interesting aspect in our DMCHA/polystyrene drop dissolution is the switching process at the drop surface by the acidic solution. DMCHA is sparsely soluble in water in neutral form, but highly soluble in protonated form. The main contribution to the dissolution is driven by the switching of DMCHA to DMCHAH$^{+}$. 

\begin{equation*}
c_s = [DMCHA^0]_{aq} + [DMCHAH^+]_{aq}
\end{equation*}
With $[DMCHA^0]_{aq}$ the natural solubility of neutral DMCHA in its environment, and $[DMCHAH^+]_{aq}$ the concentration of the protonated form of DMCHA. Here, the solubility of DMCHA can be split into two parts, and we can consider the solubility of $[DMCHA^0]_{aq}$ independent of the acid concentration.  \cite{durelle2014modelling} 


The dissolution of DMCHA is strongly dependent on the solution pH. The dissociation constant of DMCHA is expressed as 

\begin{equation}
K_a = \frac{[H^+]_{aq}[DMCHAH^0]_{aq}}{[DMCHA^+]_{aq}} 
= \frac{10^{-pH}[DMCHAH^0]_{aq}}{[DMCHA^+]_{aq}} 
\end{equation}

An increase of the acid concentration results in an increase in $[DMCHAH^+]_{aq}$ amd therefore an increase in $c_s$. The mass transfer of solvent away from the surface of the drop will be related to the concentration difference between the surface of the drop and far away from the drop $\Delta c = c_{surface} - c_{\infty}$. For a fast reaction, the equilibrium is reached and the surface concentration corresponds to the DMCHA solubility $c_s$. The concentration far away from the drop can be neglected considering the size of the drop compared to the volume of the trigger bath. Therefore, the concentration difference between the surface and far away from the drop can be reduced to the DMCHA solubility. As an increase in solubility of DMCHA at lower pH  drives faster dissolution, the lifetime $\tau$ of the drop is shorter when other conditions are fixed.

To compare the times scales of reaction and diffusion, we introduce the Damköhler number Da, defined as the ratio of the reaction rate to the diffusive mass transport rate. We assume first order kinetics in DMCHA and in $H^+$. The diffusive mass transfer rate can be derived from Fick's law. Therefore, the Damköhler number can be expressed with the equation below.

\begin{equation}
Da = \frac{k_s [DMCHA][H^+]_{aq}}{\kappa\Delta c D} 
\end{equation}

With $k_s$ the reaction rate constant of the switching reaction, [DMCHA] the concentration of DMCHA in the drop and $[H^+]_{aq}$ the concentration of protons in the aqueous phase, $\kappa$ the diffusion coefficient of DMCHA in the surrounding liquid.

According to other work \cite{wang2017extracting}, the recovery reaction of DMCHA follows a kinetic of the first order, with $k_r$ the reaction rate constant of DMCHA recovery defined as $\frac{d[DMCHA]}{dt} = k_r[DMCHA] = -k_s[DMCHAH^+][H^+]$. The slowest recovery rate studied was $k_r = 4.2 \times 10^{-4} s^{-1}$. We calculate the switching reaction rate constant  $k_s = \frac{k_r}{K_a}$, with $K_a$ the dissociation constant of DMCHA. We find with these approximations a Damköhler number $Da = 1.7 \times 10^{12}$ in the case of an aqueous phase at pH 7, suggesting that diffusion is the rate-limiting step.  Therefore in this work, we neglect the influence of the reaction on the shrinking dynamics of the drop.

\subsection{Scaling law for drop dissolution}

The simplest model that describes the dissolution process of a pure sessile drop driven by diffusion is \cite{popov2005evaporative} 

\begin{equation}
\tau = A(\theta) \frac{\rho}{\kappa \Delta c} V_0^{\frac{2}{3}}
\end{equation}
With $A(\theta)$ a geometrical constant depending on the contact angle $\theta$ (assumed constant), $\rho$ the density of the drop material, $\Delta c$ the difference between the concentration of the drop material at the drop interface and far away from the drop, and $V_0$ the initial volume of the drop. Following a diffusion mechanism, a scaling coefficient $\gamma$ of $\frac{2}{3}$ is expected \cite{epstein1950stability,popov2005evaporative,zhang2015mixed}.


Convection in the surrounding liquid may enhance the drop dissolution. \cite{dietrich2016role} A convective flow can be induced by the change in the liquid density along the surface of the dissolving drop. As a consequence of this density change, the density-altered layer is subjected to buoyancy or sinking, creating convection cells outside the drop, and enhancing the mass transport. For convection driven drop dissolution, 


\begin{equation}
\tau = A'(\theta) (\frac{\nu \rho^4}{g\beta_c\Delta c^5 D^3})^{\frac{1}{4}} V_0^{\frac{5}{12}}
\end{equation}

With $A'(\theta)$ a new geometrical constant depending on the contact angle, $\nu$ the kinematic viscosity of the surrounding liquid and g the acceleration of gravity. $\beta_c$ is the solutal expansion coefficient defined by $\beta_c = \frac{1}{\rho_b}\frac{\partial \rho}{\partial c}$, with $\rho_b$ the pure surrounding liquid density and $\frac{\partial \rho}{\partial c}$ the change in density of the surrounding liquid while a concentration $c$ of the drop liquid is dissolved in it.






We calculate the dimensionless Rayleigh number associated with the dissolution process, quantifying the ratio of the diffusion time scale over the convection time scale. 
The Rayleigh number can be defined with the following relation \cite{dietrich2016role}:

\begin{equation}
Ra = \frac{g\beta_c\Delta c (D/2)^3}{\kappa \nu}
\end{equation}

Where g is the acceleration of gravity and all the other terms are defined as for the previous equations. The term $\beta_c \Delta c$ is simplified by the expression $\frac{\Delta \rho}{\rho_{water}}$, with $\Delta \rho = \rho_{water}^{saturated} - \rho_{water}$, the difference in density between water saturated with DMCHA and the pure water density.  This term quantifies the alteration of the density of the aqueous liquid layer at the surface of the drop. Approximating the density of the mixture by the formula $\rho_{mix} = x_{DMCHA} \rho_{DMCHA} + x_{water} \rho_{water}$, we find a density difference of 2.2 $kg.m^{-3}$ at pH 7. The calculated Rayleigh number ranges from $Ra(D = 2 mm) = 3.0 \times 10^4$ for large drops to $Ra(D = 0.2 mm) = 30$ for particles at the end of the dissolution process. Both calculated Rayleigh numbers are above the threshold Rayleigh number of 12 found for dissolution of pure drops by Dietrich et al.\cite{dietrich2016role}.  

The overall scaling coefficient $\gamma$ found from the drop dissolution study is $0.53 \pm 0.07$. This coefficient is in-between the coefficient of the pure diffusion model $\gamma$ of $0.67$ and the convection-driven model $\gamma$ of $0.42$. The reason why the dissolution scaling coefficient is different than the convection dominated process may be that the transport of DMCHA across the interface is inhibited by polystyrene in the drop. Effects from the insoluble component were also observed for the dissolution of other colloidal or polymer drops in the literature. \cite{de2002solvent, okuzono2006simple, watanabe2014microfluidic, sharratt2018microfluidic} 

\subsection{Effects of initial polystyrene concentration and initial drop volume} 

The main difference observed when changing initial polystyrene concentration composition is the prefactor $\alpha$. A higher initial polystyrene concentration in the drop means that for a given initial drop volume, less DMCHA was dissolved to complete the drop dissolution process, and, assuming the dissolution rate constant, a shorter lifetime or a smaller prefactor is expected.

On the other hand, the mass transfer of DMCHA is also dependent on how much DMCHA reacts at the aqueous interface. A high initial polystyrene concentration in the drop reduces the DMCHA concentration at the drop surface. The dissolution is slowed down and an extended lifetime of the drop can be expected. Shorter lifetime at higher polystyrene concentration suggests that the main effect from polystyrene is not the reduced DMCHA dissolution rate, but less DMCHA in the drop to dissolve before the polystyrene particle formed at the end.



\subsection{In-drop phase separation and porous structure of particles} 


As the drop is immersed in the trigger solution, there is an influx of water through inter-diffusion into the drop.  Water acts as a non-solvent for polystyrene in the drop, hence reducing greatly the solubility of polystyrene in the drop and resulting in fast precipitation of polystyrene from the initial contact with the trigger solution. 
 

 During the dissolution, DMCHA flows out from the drop, and the concentration of polystyrene in the drop increases. This change in polystyrene concentration decreases the water solubility inside the drop which then leads to water microdroplets formed from the oversaturation.  These water microdroplets can exist for long time, possibly attributed to the protonated form of DMCHA acting as a surface active species. The water microdroplets contained in the main drop are trapped during the transition of the dissolving drop to a polystyrene particle. Some of the microdroplets present at the surface during this transition create a porous structure at the surface of the particles.

The movement of the water microdroplets is reminiscent of the coffee-ring effect \cite{deegan1997capillary}, recently shown to also exist in dissolving drops \cite{poulichet2020liquid}.  The large increase in viscosity of the main drop due to the increase in polystyrene concentration slows down the flow inside of the drop and fixes the water microdroplets. 

\section{Conclusions}

In this work, we studied the dynamics and in-drop phase separation during the dissolution process of drops of SHS, water and polymer in a trigger solution. Switching DMCHA to the hydrophilic form at the drop surface led to the dissolution of the drops of DMCHA and polystyrene. An increase in both trigger concentration and initial polystyrene composition decreased the lifetime of the drop. The lifetime of the dissolving drops followed a scaling relationship between the drop lifetime and the initial drop volume. An overall scaling coefficient of $0.53 \pm 0.07$ was found based on 70 different initial conditions, suggesting that the dissolution
rate was in-between the diffusion-dominated and convection driven dissolution. 
Along with drop dissolution, water intake into the drop led to in-drop precipitation of polystyrene as water acts as a non-solvent for polystyrene.  Water microdroplets formed through in-drop nucleation from reduced solubility of water in concentrated drops, and contributed to the porous microstructure of the final polystyrene particle.

A challenge related to the application of SHS such as the formation of latex \cite{su2017preparing} or extraction of bitumen \cite{holland2012separation} is how to remove SHS efficiently during switching processes.  An improved understanding of the mechanism underlying the drop dissolution, internal drop dynamics and final structure formation may be useful for rational design of the switching procedures for reduced solvent residue and desirable properties of the final particles.

\section*{Acknowledgements}

We acknowledge Mitacs Accelerate Program, BC Research,the Canada Foundation for Innovation (CFI), the Natural Science and Engineering Research Council of Canada (NSERC) and Alberta-innovates for funding support. This work is partially supported by the Canada Research Chairs program. We are grateful for the inspiring discussions with Dr Hassan Hamza and the co-authors at BC Research, and for the technical assistance from IOSI Labs at Faculty of Engineering.

\section*{Conflicts of interest}

BC Research, is a wholly owned subsidiary of NORAM Engineering and Constructors Ltd., which licensed the SHS technology platform originally developed and Patented at Queens University and are together developing new commercial applications for the SHS technology.

\bibliographystyle{unsrt}  
\bibliography{references}  

\end{document}